\documentclass[useAMS,usenatbib]{mn2e}
\newcommand{\xmm}{{\it XMM-Newton}}

\newcommand{\rxj}{\hbox{{RX\,J0720.4$-$3125}}}

\def\gtrsim{\mathrel{\hbox{\rlap{\hbox{\lower4pt\hbox{$\sim$}}}\hbox{$>$}}}}
\usepackage{graphicx}  
\usepackage{longtable}                
\usepackage{subfig}
\usepackage{caption}
\usepackage{everysel}
\usepackage{keyval}
\usepackage{ragged2e}
\usepackage{rotating}
\usepackage[draft,pdftex,pdfpagemode={UseOutlines},bookmarks,bookmarksopen,colorlinks,linkcolor={blue},citecolor={green},urlcolor={red}]{hyperref}
\bibpunct{(}{)}{;}{a}{}{,}

\title[Spectral evolution of \rxj{}.]{The continued spectral and temporal evolution of \rxj{}}
\author[M. M. Hohle et al.]{M. M. Hohle$^{1}$, F. Haberl$^{2}$, J. Vink$^{3}$, C.P. de Vries$^{4}$, R. Turolla$^{5,6}$, S. Zane$^{6}$, \newauthor and M. M\'endez$^{7}$\\
$^{1}$Astrophysikalisches Institut und Universit\"ats-Sternwarte Jena, Schillerg\"asschen 2-3, 07745 Jena, Germany; mhohle@astro.uni-jena.de\\
$^{2}$Max-Planck-Institut f\"ur extraterrestrische Physik, Giessenbachstra{\ss}e, 85741 Garching, Germany\\
$^{3}$Astronomical Institute Anton Pannekoek, University of Amsterdam, P.O. Box 94249 1090 GE Amsterdam, The Netherlands\\
$^{4}$SRON, Netherlands Institute of Space Research, Sorbonnelaan 2, 3584 CA, Utrecht, The Netherlands\\
$^{5}$Department of Physics and Astronomy, University of Padua, via Marzolo 8, 35131 Padova, Italy\\ 
$^{6}$Mullard Space Science Laboratory, University College London, Holmbury St. Mary, Dorking, Surrey, RH5, 6NT, UK\\
$^{7}$Kapteyn Astronomical Institute, University of Groningen, P.O. Box 800, 9700 AV Groningen, The Netherlands}

\begin{document}

\bibliographystyle{mn2e}

\date{Accepted 201X Month XX. Received 201X Month XX; in original form 201X Month XX}

\pagerange{\pageref{firstpage}--\pageref{lastpage}} \pubyear{2002}

\maketitle

\label{firstpage}

\begin{abstract}
\rxj{} is the most peculiar object among a group of seven isolated X-ray pulsars (the so-called ``Magnificent Seven"), since it shows long-term variations of its spectral and temporal properties on time scales of years. This behaviour was explained by different authors either by free precession (with a seven or fourteen years period) or possibly a glitch that occurred around $\mathrm{MJD=52866\pm73~days}$.\\
We analysed our most recent \xmm{} and {\it Chandra} observations in order to further monitor the behaviour of this neutron star. With the new data sets, the timing behaviour of \rxj{} suggests a single (sudden) event (e.g. a glitch) rather than a cyclic pattern as expected by free precession. The spectral parameters changed significantly around the proposed glitch time, but more gradual variations occurred already before the (putative) event. Since $\mathrm{MJD\approx53000~days}$ the spectra indicate a very slow cooling by $\sim$2~eV over 7 years.

\end{abstract}


\section{Introduction}
\begin{figure*}
\centering
\includegraphics*[viewport=95 230 500 665, width=0.825\textwidth]{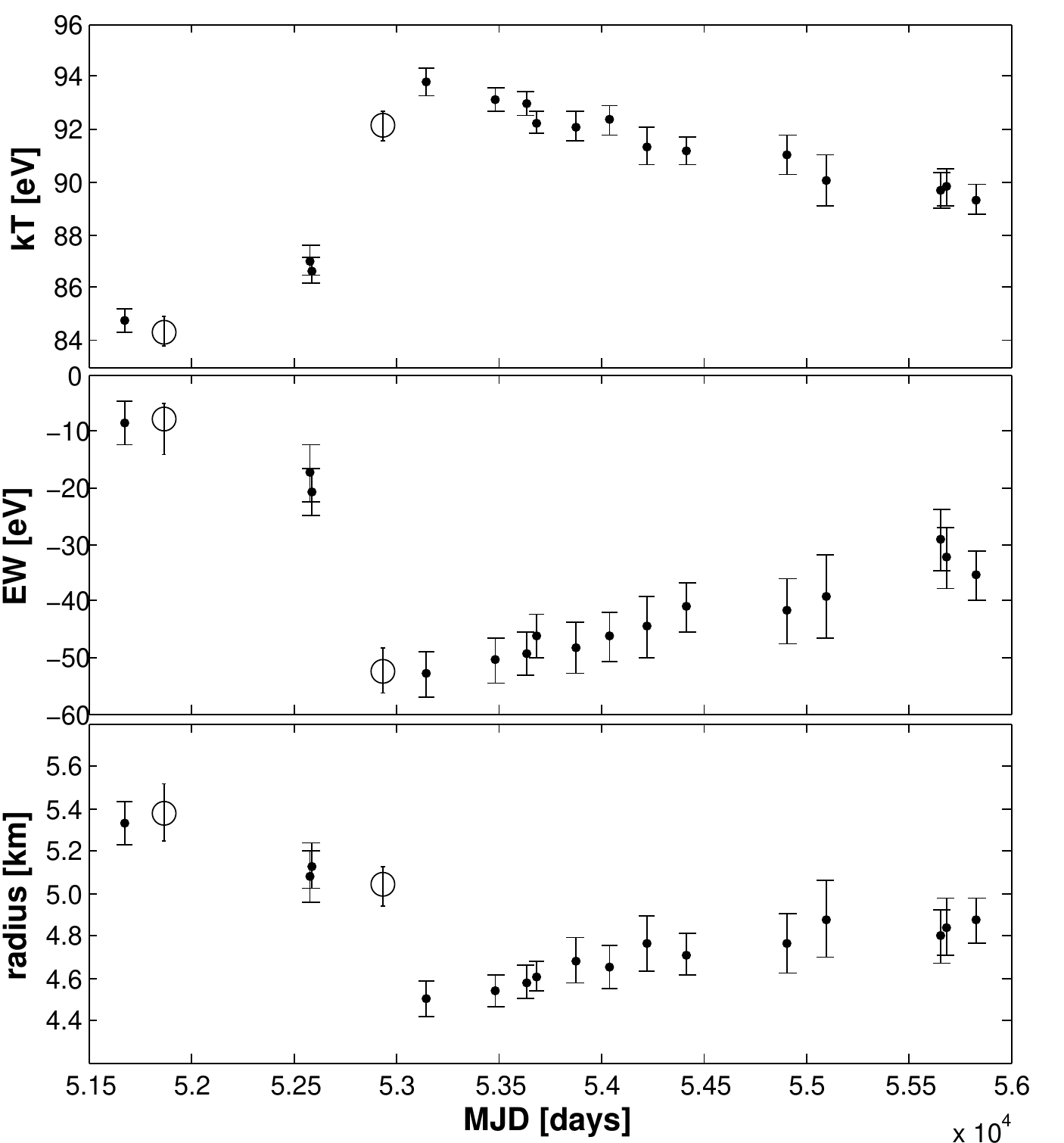}
\caption[]{Spectral properties of RX\,J0720, as listed in \autoref{ffthin_0720_tab}.}
\label{kTEWR}%
\end{figure*}
The isolated neutron star (NS) \rxj{} (RX\,J0720) belongs to a group of seven nearby ($\mathrm{\la500~pc}$) radio-quiet X-ray pulsars, the so-called ``Magnificent Seven" (M7), discovered as bright X-ray sources in the ROSAT all-sky survey data. The M7 exhibit soft ($\mathrm{T_{eff}\approx40-100~eV}$) blackbody-like X-ray spectra, in some cases with one or more broad absorption features that are interpreted as proton-cyclotron resonances or atomic transitions of bound species in a strong magnetic field, $\mathrm{B\approx10^{13}-10^{14}~G}$. Assuming magnetic dipole braking, similar magnetic field strengths can be derived from the standard spin-down formula for those sources for which pulse periods (all in the 3-12~s interval) and period derivatives are measured (see \citealt{2009ApJ...705..798K,2009ApJ...692L..62K,2008ApJ...673L.163V}). The M7 have ages of $\mathrm{0.3-3~Myrs}$, as inferred by cooling curves or kinematics \citep{2010MNRAS.402.2369T,2002ApJ...571..447K}, while the characteristic ages are larger ($\mathrm{2-5~Myrs}$). For a detailed review of the M7, we refer to \citet{2007Ap&SS.308..181H} and \citet{2009ApJ...705..798K}.\\
RX\,J0720 is the second brightest member of the M7 and it was identified as a pulsating X-ray source with a 8.39~s spin period in \citet{1997A&A...326..662H}. \citet{2001A&A...365L.302C} discovered a hardness ratio variation with pulse phase and a phase shift between the flux and the hardness ratio in the \xmm{} data of RX\,J0720. Based on \xmm{} RGS data, \citet{2004A&A...415L..31D} showed that the energy-dependent change in the pulse profile is accompanied by a long term change of the X-ray spectrum\footnote{Note that initially the changes at long wavelengths were overestimated, as the \xmm{} RGS instrument suffers from a decline in sensitivity in the long wavelength band.}. The spectral changes were soon confirmed using \xmm{} EPIC \citep{2004A&A...419.1077H} and {\it Chandra} LETG-S data \citep{2004ApJ...609L..75V}. Furthermore, \citet{2006A&A...451L..17H} found a phase lag between soft ($\mathrm{0.12-0.40~keV}$) and hard ($\mathrm{0.40-1.00~keV}$) photons, which changes over years. The \xmm{} spectra of RX\,J0720 are best modelled with a blackbody plus a broad absorption feature at $\mathrm{\sim0.3~keV}$ \citep{2004A&A...419.1077H}. \citet{2006A&A...451L..17H} reported variations of the blackbody temperature, the equivalent width of the absorption feature and the blackbody normalisation of RX\,J0720 compatible with a periodic behaviour with a long term period of $\mathrm{P_{long}\approx7.1~yrs}$. However, the data used in \citet{2006A&A...451L..17H} spanned only 4.5~yrs, i.e. not the complete cycle of the tentative period.\\
\begin{table}
\centering
\caption[]{The three new \xmm{} observations (all with thin filter) performed after \citet{2010A&A...521A..11H}. We list the net counts in soft band (0.12-0.40~keV) and hard band (0.40-1.00~keV). The soft photons from the EPIC-MOS data are not used in this work.}
\label{new_0720_tab}
\begin{tabular}{llrcc}
\hline
MJD [days]/	 & EPIC	    &	eff exp	& net cts & net cts \\
obsID        & setup 		& [ks]   	& soft		& hard \\		
\hline
55662 /      & pn   /FF  & 14.41		& 53014			& 38114 \\
0650920101   & MOS1 /SW  & 20.47		& --			  & 10429 \\
             & MOS2 /SW  & 19.86		& --			  & 10260 \\
             \hline
55684/       & pn   /FF  & 13.10		& 48680			& 35436 \\
0670700201   & MOS1 /SW  & 22.02		& --			  & 11934 \\
             & MOS2 /SW  & 23.20		& --			  & 12224 \\
             \hline
55835/       & pn   /FF  & 22.18    & 86713			& 58888 \\
0670700301   & MOS1 /SW  & 25.82		& --			  & 12975 \\
             & MOS2 /SW  & 25.83		& --			  & 13042 \\
\hline
\end{tabular}
\end{table}
The period derivative of RX\,J0720 was first estimated by \citet{2002MNRAS.334..345Z} and subsequently further constrained by \citet{2004MNRAS.351.1099C} and \citet{2005ApJ...628L..45K} as new observations become available. \citet{2006A&A...451L..17H} found that periodical phase residuals were possibly present (again with $\mathrm{P_{long}\approx7.5~yrs}$) in the timing solution of RX\,J0720 with a constant value of $\dot P=0.698(2)\times10^{-13}~s~s^{-1}$ \citep{2005ApJ...628L..45K}.\\
The spectral and temporal variations of RX\,J0720 are unique among the M7 and were explained either by free precession \citep{2006A&A...451L..17H,2007Ap&SS.308..181H,2009A&A...498..811H}, or a glitch that occurred at $\mathrm{MJD=52866\pm73~days}$ \citep{2007ApJ...659L.149V}. Both scenarios, free precession and the glitch event, have their drawbacks, as discussed in \citet{2007ApJ...659L.149V} and \citet{2009A&A...498..811H,2010A&A...521A..11H}.\\ 
The most recent overview of the spectral evolution of RX\,J0720 was given in \citet{2009A&A...498..811H}. Since then, our team performed five further \xmm{} observations. Moreover, three {\it Chandra} observations were obtained after the last update of the timing solution \citep{2010A&A...521A..11H}. Here, we present our analysis and results for these more recent data sets in connection with further spectral and temporal evolution of RX\,J0720. 
\section[]{Data and data reduction}
In addition to \citet{2009A&A...498..811H} we analyse here five new \xmm{} observations, two of which (revolutions 1700 and 1792) were already used for the timing in \citet{2010A&A...521A..11H}, but not to investigate the spectral behaviour. We reduced all available \xmm{} data of RX\,J0720 with the standard \xmm\ {\bf S}cience {\bf A}nalysis {\bf S}ystem (SAS) version 11.0 using the {\sc epchain} and {\sc emchain} tasks for EPIC-pn \citep{2001A&A...365L..18S} and both EPIC-MOS \citep{2001A&A...365L..27T}, respectively. For details on the analysis of the \xmm{} data (i.e. data extraction and good time interval, GTI, filtering) we refer to \citet{2009A&A...498..811H,2010A&A...521A..11H,2012MNRAS.419.1525H}. We list the three new data sets (neither used for timing, nor for spectroscopy so far) in \autoref{new_0720_tab}. EPIC-MOS was always used in small window (SW) mode with a time resolution of 0.3~s, whereas EPIC-pn was used in full frame mode (FF, time resolution of 73.4~ms).\\
We analysed the {\it Chandra} HRC-S/LETG \citep{1996ChNew...4....9J} data with CIAO 4.1 and refer to \citet{2010A&A...521A..11H,2012MNRAS.419.1525H} for details, both on the data reduction and the most recent {\it Chandra} HRC-S/LETG observations (SRON and MPE guaranteed time data) of RX\,J0720.\\ Due to the lack of a sufficient amount of photons for the individual spectra, we use the {\it Chandra} data for timing analysis only.
\begin{figure*}
\centering
\includegraphics*[viewport=65 4 477 719, width=0.625\textwidth]{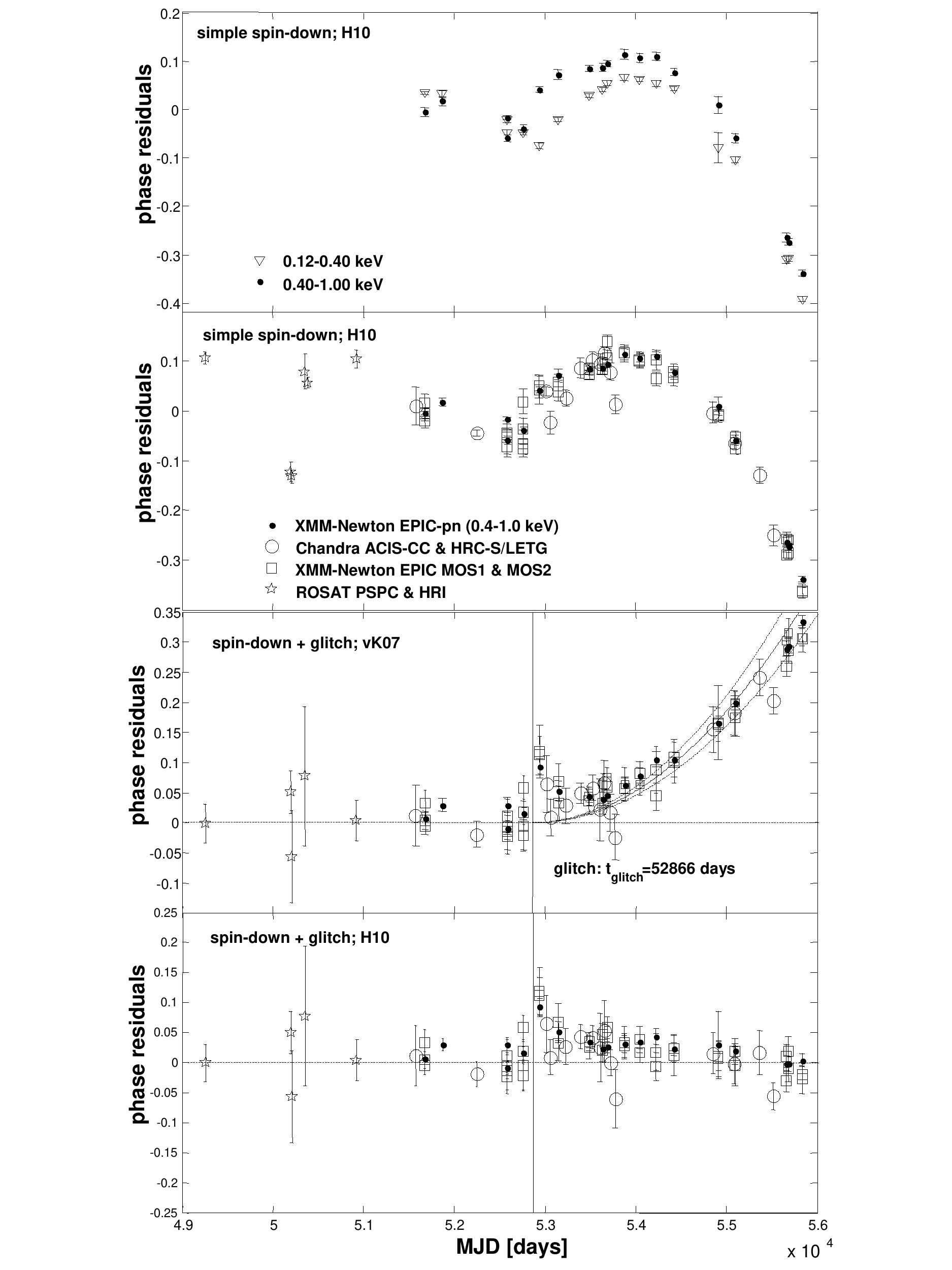}
\caption[]{Phase residuals of RX\,J0720 after applying the phase coherent ``all data" timing solution in \citet[H10]{2010A&A...521A..11H} with constant spin-down (upper two panels). The top panel illustrates the variable phase shift between soft (0.12-0.40~keV) and hard (0.40-1.00~keV) photons seen in the EPIC-pn data. The glitch solution proposed by \citet[vK07]{2007ApJ...659L.149V} fits well the data available at this time ($\mathrm{MJD=53500~days}$), but poorly represents the data at later epochs (third panel). The later (deviant) points require an additional quadratic term (shown with its $\mathrm{1\sigma}$ uncertainty) to be explained; this corresponds to a modification of the spin-down parameter $\mathrm{\dot{f}}$ for $\mathrm{t>t_{g}}$ \citep{2010A&A...521A..11H}. The phase residuals from the most recent observations (after $\mathrm{MJD=55100~days}$) are consistent with the modified glitch solution (lower panel). The glitch time $\mathrm{t_{g}}$ at $\mathrm{MJD=52866\pm73~days}$ is indicated by the solid vertical line in the lower two panels. All error bars correspond to $\mathrm{1\sigma}$ confidence.}
\label{residuals}%
\end{figure*}
\section[]{Results}

\subsection[]{Spectral behaviour}
To investigate the spectral evolution of RX\,J0720 we first fitted all sixteen EPIC-pn spectra obtained in full frame mode with thin filter in one session using {\sc Xspec12}. We used the model {\it phabs*(bbodyrad+gaussian)}, as also used in \citet{2004A&A...419.1077H,2006A&A...451L..17H,2007Ap&SS.308..181H,2009A&A...498..811H} and \citet{2012MNRAS.419.1525H}, where {\it gaussian} is used to account for the broad absorption feature at 0.3~keV. The simultaneous fit of the sixteen EPIC-pn datasets results in $\mathrm{\chi^2/d.o.f=1.23}$ with 2374 degrees of freedom. We obtain $\mathrm{N_H=0.984\pm0.050\times10^{20}~cm^{-2}}$ for interstellar absorption, $\mathrm{E_{line}=311.9\pm5.0~eV}$ for the central energy and $\mathrm{\sigma=64.4\pm3.5~eV}$ for the line width of the broad absorption feature (all errors denote 90\% confidence level). These three parameters were assumed to be constant for all observations, as in previous works \citep{2004A&A...419.1077H,2006A&A...451L..17H,2007Ap&SS.308..181H,2009A&A...498..811H}. The blackbody temperature (kT), emitting radius (R, computed assuming a distance of $\mathrm{D=300~pc}$, see \citealt{2007ApJ...660.1428K} and \citealt{EisenPhd}), and the line equivalent width (EW) were allowed to vary between the observations.\\ 
Due to cross-calibration and pile-up issues for the normalisation (see \citealt{2004A&A...419.1077H} for a detailed discussion), the data obtained with other instrument setups (revolutions 0175 with medium filter and revolution 0711 in small window mode and medium filter) were fitted separately, but fixing $\mathrm{N_H}$, $\mathrm{E_{line}}$ and $\sigma$ at the values obtained from the simultaneous fit of the sixteen EPIC-pn spectra performed in full frame mode with thin filter, see \autoref{ffthin_0720_tab} and \autoref{kTEWR}.\\
As reported already in \citet{2006A&A...451L..17H}, the temperature, size of the emitting area and equivalent width underwent major changes around $\mathrm{MJD=53000~days}$, but since then (i.e. over the last seven years), all three parameters changed only gradually. \xmm{} observations cover now a time span of almost 12~yrs, hence a 7.5~yrs period can be excluded. A 14~yrs period seems unlikely, since if one extrapolates the spectral evolution (\autoref{kTEWR}) for two further years, the spectral properties are still significantly different to their initial values.
\begin{table*}
\centering
\caption[]{Effective temperature (kT) and radiation radius (both measured at infinity), flux and equivalent width (EW) of the broad absorption feature derived from \xmm{} EPIC-pn spectra (see also \autoref{kTEWR}). All observations were performed in full frame mode with thin filter, except rev. 0175 (medium filter) and rev. 0711 (small window mode with medium filter), that are highlighted in italic. All errors denote 90\% confidence level.}
\label{ffthin_0720_tab}
\begin{tabular}{lccccc}
\hline
ReNo&	MJD &	kT	&	radius	& EW  & flux 0.12-1.0~keV \\
    &  [days]& 	[eV]   	& [km]		& [eV] & [$\mathrm{10^{-11}ergs~cm^{-2}~s^{-1}}$]\\		
\hline
0078   &51677  & $\mathrm{84.74\pm0.43}$&  $\mathrm{5.33_{-0.10}^{+0.10}}$ &$\mathrm{ -8.9_{-3.9}^{+3.8}}$  & $\mathrm{1.0903_{-0.0100}^{+0.0094}}$   \\
{\it 0175} & {\it 51870} &  $\mathrm{84.32\pm0.57}$ & $\mathrm{5.38_{-0.14}^{+0.13}}$ & $\mathrm{ -7.9_{-6.1}^{+2.9}}$ &  $\mathrm{1.1061_{-0.0062}^{+0.0062}}$ \\	 
0533   &52585  & $\mathrm{87.01\pm0.57}$&  $\mathrm{5.08_{-0.12}^{+0.12}}$ &$\mathrm{-17.2_{-5.1}^{+5.0}}$  & $\mathrm{1.0914_{-0.0108}^{+0.0092}}$   \\
0534   &52587  & $\mathrm{86.63\pm0.48}$&  $\mathrm{5.13_{-0.10}^{+0.10}}$ &$\mathrm{-20.7_{-4.3}^{+4.0}}$  & $\mathrm{1.0800_{-0.0080}^{+0.0060}}$   \\
{\it 0711} & {\it 52940} & $\mathrm{92.12\pm0.56}$&   $\mathrm{5.04_{-0.11}^{+0.10}}$ & $\mathrm{-52.3_{-3.8}^{+4.0}}$ & $\mathrm{1.2638_{-0.0096}^{+0.0064}}$\\	
0815   &53148  & $\mathrm{93.80\pm0.51}$&  $\mathrm{4.50_{-0.09}^{+0.08}}$ &$\mathrm{-52.8_{-4.1}^{+4.0}}$  & $\mathrm{1.1007_{-0.0074}^{+0.0066}}$   \\
0986   &53489  & $\mathrm{93.14\pm0.45}$&  $\mathrm{4.54_{-0.08}^{+0.08}}$ &$\mathrm{-50.5_{-4.0}^{+4.0}}$  & $\mathrm{1.0883_{-0.0066}^{+0.0075}}$   \\
1060   &53636  & $\mathrm{92.97\pm0.47}$&  $\mathrm{4.58_{-0.08}^{+0.08}}$ &$\mathrm{-49.2_{-3.9}^{+3.8}}$  & $\mathrm{1.1028_{-0.0055}^{+0.0064}}$   \\
1086   &53687  & $\mathrm{92.24\pm0.41}$&  $\mathrm{4.61_{-0.07}^{+0.07}}$ &$\mathrm{-46.2_{-3.9}^{+3.8}}$  & $\mathrm{1.0860_{-0.0080}^{+0.0061}}$   \\
1181   &53877  & $\mathrm{92.11\pm0.59}$&  $\mathrm{4.68_{-0.11}^{+0.10}}$ &$\mathrm{-48.2_{-4.6}^{+4.5}}$  & $\mathrm{1.1061_{-0.0083}^{+0.0039}}$   \\
1265   &54045  & $\mathrm{92.35\pm0.58}$&  $\mathrm{4.65_{-0.10}^{+0.10}}$ &$\mathrm{-46.1_{-4.7}^{+4.2}}$  & $\mathrm{1.1130_{-0.0100}^{+0.0080}}$   \\
1356   &54226  & $\mathrm{91.35\pm0.71}$&  $\mathrm{4.76_{-0.13}^{+0.13}}$ &$\mathrm{-44.4_{-5.7}^{+5.3}}$  & $\mathrm{1.1110_{-0.0120}^{+0.0040}}$   \\
1454   &54421  & $\mathrm{91.17\pm0.54}$&  $\mathrm{4.71_{-0.10}^{+0.10}}$ &$\mathrm{-40.9_{-4.6}^{+4.1}}$  & $\mathrm{1.0900_{-0.0079}^{+0.0060}}$   \\
\multicolumn{6}{c}{observations since \citet{2009A&A...498..811H}}\\
1700   &54912  & $\mathrm{91.05\pm0.75}$&  $\mathrm{4.76_{-0.14}^{+0.14}}$ &$\mathrm{-41.6_{-5.9}^{+5.6}}$  & $\mathrm{1.1042_{-0.0120}^{+0.0096}}$   \\
1792   &55096  & $\mathrm{90.08\pm0.96}$&  $\mathrm{4.87_{-0.18}^{+0.18}}$ &$\mathrm{-39.1_{-7.3}^{+7.1}}$  & $\mathrm{1.1067_{-0.0140}^{+0.0110}}$   \\
2076   &55662  & $\mathrm{89.69\pm0.65}$&  $\mathrm{4.80_{-0.13}^{+0.12}}$ &$\mathrm{-29.1_{-5.5}^{+5.2}}$  & $\mathrm{1.0841_{-0.0120}^{+0.0098}}$   \\
2087   &55684  & $\mathrm{89.80\pm0.71}$&  $\mathrm{4.84_{-0.14}^{+0.13}}$ &$\mathrm{-32.4_{-5.5}^{+5.2}}$  & $\mathrm{1.0990_{-0.0081}^{+0.0040}}$   \\
2163   &55835  & $\mathrm{89.34\pm0.54}$&  $\mathrm{4.87_{-0.11}^{+0.10}}$ &$\mathrm{-35.3_{-4.5}^{+4.1}}$  & $\mathrm{1.0776_{-0.0110}^{+0.0063}}$   \\   
\hline
\end{tabular}
\end{table*}
\subsection[]{Timing}

Applying the ``all data" solution with constant spin-down, \citet{2010A&A...521A..11H} found that the phase residuals of RX\,J0720 have shown a long term behaviour with a possible periodic pattern yielding a 7-9~yr or a 14-16~yr period (depending on assumptions) until summer 2010 ($\mathrm{MJD\approx55400~days}$). Therefore, it was expected that the phase residuals (which were negative at the time of the previous investigation) will approach zero for the next observations. However, the phase residuals still reach large and negative values, if the ``all data" solution in \citet{2010A&A...521A..11H} is applied (\autoref{residuals}, upper two panels) to the new data. Also the variable phase shift between soft and hard photons (\autoref{residuals}, upper panel) stays constant since the last observations, whereas it was expected that the phase shift will reverse sign again, like it occurred around $\mathrm{MJD=53000~days}$, if RX\,J0720 precesses.\\  
\citet{2007ApJ...659L.149V} proposed a ``glitch solution" to explain the timing behaviour of RX\,J0720, that well fits the data available at that time (see \autoref{residuals}, third panel), but does not represent the data after $\mathrm{MJD=53500~days}$. \citet{2010A&A...521A..11H} modified the ``glitch solution" of \citet{2007ApJ...659L.149V} by including a change in spin-down $\mathrm{\dot f_{c}}$, valid for $\mathrm{t>t_{g}}$ (\autoref{glitch_tab}). This term corrects the drift of the phase residuals in \autoref{residuals} (third panel), since the time span available for \citet{2007ApJ...659L.149V} was too short for a more accurate extrapolation of the phase. Including $\mathrm{\dot f_{c}}$, even the phase residuals of the data that were not available to \citet[i.e., after $\mathrm{MJD=55100~days}$]{2010A&A...521A..11H} are consistent with zero (\autoref{residuals}, lowest panel). Hence, the ``glitch solution" of \citet{2007ApJ...659L.149V} with the update of \citet{2010A&A...521A..11H} models the timing behaviour of RX\,J0720 much better than a timing solution with constant spin-down.

\begin{table}
\centering
\caption[]{The timing parameters of the glitch solution \citep{2007ApJ...659L.149V} for RX\,J0720 with $\mathrm{\dot f_{c}}$ for $\mathrm{t>t_g}$. The numbers in parenthesis indicate the $\mathrm{2\sigma}$ errors. The phase is determined by $\mathrm{\Phi(t)=\Phi(t_o)+f(t-t_o)+0.5\cdot\dot f (t-t_o)^2 - 0.5\cdot\dot f_c (t-t_o)^2+ \Delta\Phi_g(t)}$, with $\mathrm{\Delta\Phi_g(t)=-\Delta f(t-t_o)-0.5\cdot\Delta\dot f (t-t_o)^2}$ and $\mathrm{\Delta\Phi_g(t)=0}$ for $\mathrm{t>t_g}$.}
\label{glitch_tab}
\begin{tabular}{lc}
		
\hline

$\mathrm{t_{0}}$ [MJD]     				&53,010.2635667(10)\\
f [Hz] 							&0.1191736716(9)\\
$\mathrm{\dot f}$ [$\mathrm{10^{-15}~Hz/s}$]		&-1.04(3)\\
$\mathrm{t_{g}}$ [MJD]						&52,866(73)\\
$\mathrm{\Delta f}$ [nHz]				&4.1(12)\\
$\mathrm{\Delta \dot f}$ [$\mathrm{10^{-17}~Hz/s}$]	&-4(3)\\
$\mathrm{\dot f_{c}}$ [$\mathrm{10^{-17}~Hz/s}$]	&-1.11(20)\\
\hline
\end{tabular}
\end{table}

\begin{figure}
\centering
\includegraphics*[viewport=100 275 490 610, width=0.425\textwidth]{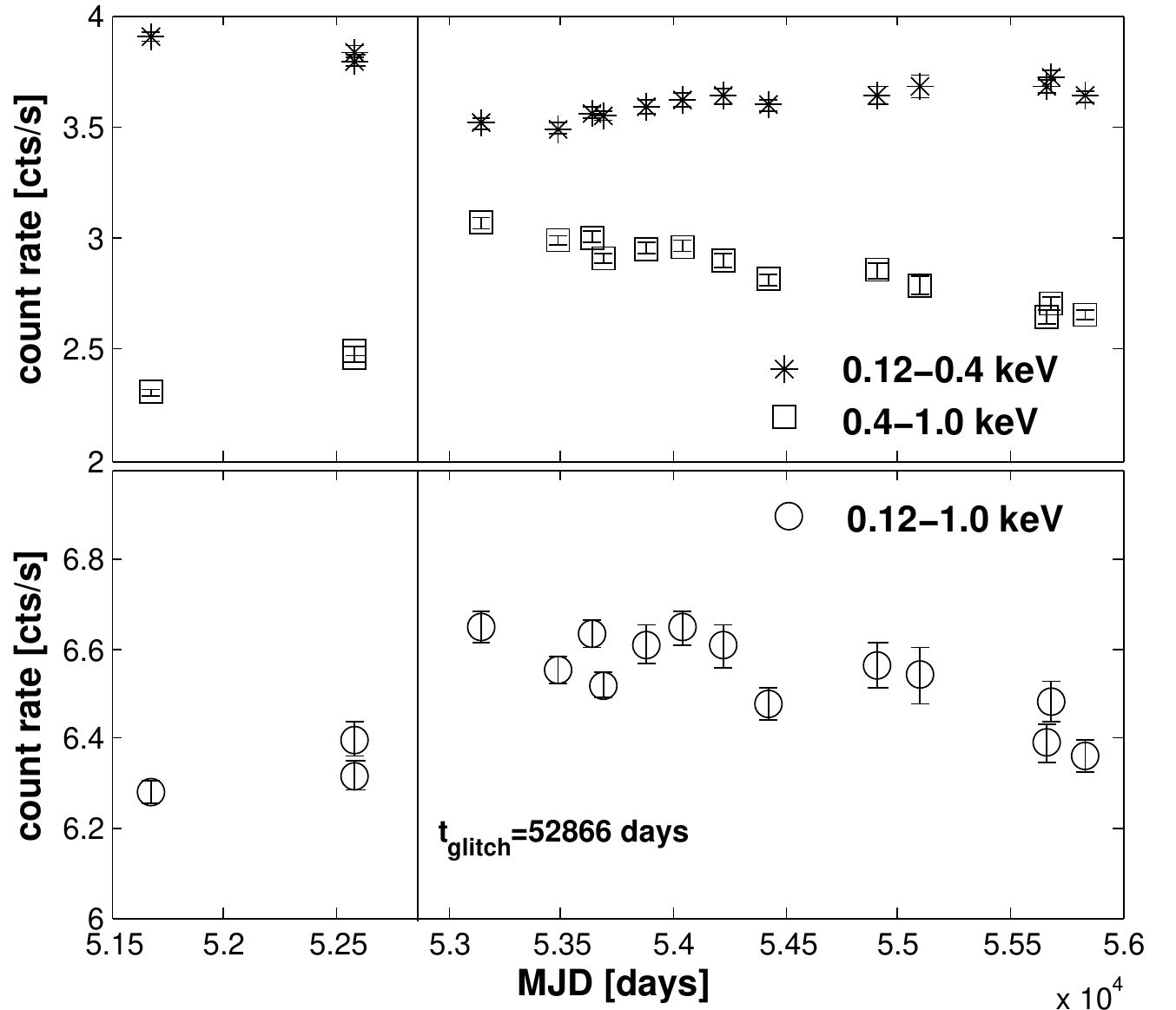}
\caption[]{The evolution of the count rates of RX\,J0720 in the soft and hard band (upper panel) and the total count rate (lower panel). The data are obtained for the observations performed with \xmm{} EPIC-pn in full frame mode with thin filter (errors denote 90\% confidence level).}
\label{cts}%
\end{figure}

\section{Discussion}
\label{discussion}
The present data of RX\,J0720 do not support a cyclic behaviour with a period in the 7 to 14 year range in the spectral and timing properties of the source. However, the measured blackbody temperature is still declining and, by extrapolating the linear trend (since the proposed glitch time $\mathrm{t_g=52866\pm73~days}$), RX\,J0720 will reach its initial state in autumn 2019. It is of interest to follow this decline and assessing whether the temperature will finally stabilise at the pre-2003 value. This requires a monitoring for at least 20 years in total (out of which eleven years have been already covered) to reveal any long term periodicity. It is not possible to explain both, the spectral and temporal changes of RX\,J0720 by precession with a self-consistent model similar to that discussed in \citet{2006A&A...451L..17H}, figure~5 therein. This might reflect the lack of knowledge regarding to the exact emission geometry (spot shape, temperature distribution, atmospheric effects etc.) of this NS.\\
The `glitch solution" of \citet{2007ApJ...659L.149V} fits well the phase residuals, if the modification by \citet{2010A&A...521A..11H} is applied. The jump in frequency at $\mathrm{MJD=52866\pm73~days}$ would correspond to the gain of angular momentum imparted by a mass of $\mathrm{10^{20}-10^{21}~g}$ accreted by the NS \citep{2007ApJ...659L.149V}. Hence, the glitch might have been caused by an accretion event e.g. the impact of an asteroid. Recently, some evidence for a disc or a dense ($\mathrm{n_{H}=10-10^{10}~cm^{-3}}$) ambient medium around RX\,J0720 was discussed \citep{2009A&A...497L...9H,2012MNRAS.419.1525H}. This (still) hypothetic disc may host material for such an impact (see discussion in \citealt{2012MNRAS.419.1525H}). However, as illustrated in \autoref{kTEWR}, the spectral changes occurred already before $\mathrm{MJD=52866\pm73~days}$ and this would point rather to a slow change than a sudden event. Also, the variable phase lag between soft and hard photons is difficult to reconcile with an impact. Moreover, the total flux (120$-$1000~eV, \autoref{ffthin_0720_tab}) of RX\,J0720 remained almost constant, but the fluxes in the soft and the hard band changed significantly (the spectrum became harder  until $\mathrm{MJD\approx53000~days}$ and now it is softening again, see \autoref{cts}, suggesting the existence of at least two emission regions with different temperatures). The best fit blackbody temperature and size of the emitting area show changes of $\mathrm{\approx10-20\%}$. However, it is remarkable that these changes somehow conspire to keep the flux within 10\% variation, showing that the changes cannot be caused by a sudden heating alone.\\
In the case of a glitch or an impact, the total flux is expected to increase. The changes indicate a re-arranging of the flux, rather than heating by a glitch. The long-term changes in the absorption feature are also suggestive of some gradual, non-impulsive mechanism behind the timing behaviour of the source. This leads to the conclusion, that the spectral and temporal evolution of RX\,J0720 might be caused by magnetospheric distortions, hence a re-arranging of the magnetic field. Note, that the broad band luminosity remains constant in the X-rays, but the source was not monitored at optical and UV wavelengths \citep{1998A&A...333L..59M,2003A&A...408..323M,2003ApJ...590.1008K,2010AN....331..243E}.\\
RX\,J0720 is close to magnetars in the $\mathrm{P-\dot{P}}$ diagram. In particular, evolutionary connections between the M7 and the soft gamma ray repeaters (SGRs) and anomalous X-ray pulsars (AXPs) have been discussed by many authors \citep{1998ApJ...506L..61H,2009ApJ...692L..62K,2010MNRAS.401.2675P}. AXPs and SGRs have similar pulse periods and are younger (by a factor of $\mathrm{10-100}$) than the M7. Despite low statistics (only $\approx10$ objects in each group are observed) and the unsettled properties of some objects, \citet{2010MNRAS.401.2675P} have shown that the different families of NSs can be explained by one evolutionary model. A possibility, then, is that the M7 descend from SGRs/AXPs, and are aged magnetars in which the magnetic dipole field decayed from the initial $\mathrm{\approx 10^{14}~G}$ to the present $\mathrm{\approx 10^{13}~G}$. The decay of the surface field is actually triggered by the decay of the internal, toroidal and poloidal, one. It is the progressive exhaustion of internal magnetic helicity that is responsible for the low-level activity of old magnetars (bursting/outbursting behaviour, non-thermal X-ray spectral components). The recent discovery of a low-field SGR \citep{2010Sci...330..944R} and its likely interpretation as an aged magnetar \citep{2011ApJ...740..105T} lends further support to this picture. It could be that, contrary to SGR\,0418+5729, the initial internal toroidal field in RX\,J0720 was not strong enough to power an outburst in its late stages of evolution \citep{2011ApJ...727L..51P} and the last hiccups of activity are seen as moderate changes in the spectral and timing properties. In this case, we should witness more erratic spectral and temporal irregularities in the future, if the monitoring of RX\,J0720 will be continued.

\section*{Acknowledgments}

The \xmm\ project is supported by the Bundesministerium f\"ur Wirtschaft und
Technologie/Deutsches Zentrum f\"ur Luft- und Raumfahrt (BMWI/DLR, FKZ 50 OX 0001)
and the Max-Planck Society. MMH acknowledges support by the Deutsche Forschungsgemeinschaft (DFG) through
SFB/TR 7 ``Gravitationswellenastronomie''. The work of RT is funded by INAF through the PRIN 2011 grant scheme. We are grateful to SRON and MPE for allowing us to observe RX\,J0720 as part of their {\it Chandra} guaranteed time program.

\bibliography{references}

\begin{thebibliography}{}

\bibitem[\protect\citeauthoryear{{Cropper}, {Haberl}, {Zane} \&
  {Zavlin}}{{Cropper} et~al.}{2004}]{2004MNRAS.351.1099C}
{Cropper} M.,  {Haberl} F.,  {Zane} S.,    {Zavlin} V.~E.,  2004, \mnras, 351,
  1099

\bibitem[\protect\citeauthoryear{{Cropper}, {Zane}, {Ramsay}, {Haberl} \&
  {Motch}}{{Cropper} et~al.}{2001}]{2001A&A...365L.302C}
{Cropper} M.,  {Zane} S.,  {Ramsay} G.,  {Haberl} F.,    {Motch} C.,  2001,
  \aap, 365, L302

\bibitem[\protect\citeauthoryear{{de Vries}, {Vink}, {M{\'e}ndez} \&
  {Verbunt}}{{de Vries} et~al.}{2004}]{2004A&A...415L..31D}
{de Vries} C.~P.,  {Vink} J.,  {M{\'e}ndez} M.,    {Verbunt} F.,  2004, \aap,
  415, L31

\bibitem[\protect\citeauthoryear{{Eisenbeiss}}{{Eisenbeiss}}{2011}]{EisenPhd}
{Eisenbeiss} T.,  2011, PhD Thesis, University Jena, Germany

\bibitem[\protect\citeauthoryear{{Eisenbeiss}, {Ginski}, {Hohle}, {Hambaryan},
  {Neuh{\"a}user} \& {Schmidt}}{{Eisenbeiss}
  et~al.}{2010}]{2010AN....331..243E}
{Eisenbeiss} T.,  {Ginski} C.,  {Hohle} M.~M.,  {Hambaryan} V.~V.,
  {Neuh{\"a}user} R.,    {Schmidt} T.~O.~B.,  2010, Astronomische Nachrichten,
  331, 243

\bibitem[\protect\citeauthoryear{{Haberl}}{{Haberl}}{2007}]{2007Ap&SS.308..181H}
{Haberl} F.,  2007, \apss, 308, 181

\bibitem[\protect\citeauthoryear{{Haberl}, {Motch}, {Buckley}, {Zickgraf} \&
  {Pietsch}}{{Haberl} et~al.}{1997}]{1997A&A...326..662H}
{Haberl} F.,  {Motch} C.,  {Buckley} D.~A.~H.,  {Zickgraf} F.,    {Pietsch} W.,
   1997, \aap, 326, 662

\bibitem[\protect\citeauthoryear{{Haberl}, {Turolla}, {de Vries}, {Zane},
  {Vink}, {M{\'e}ndez} \& {Verbunt}}{{Haberl}
  et~al.}{2006}]{2006A&A...451L..17H}
{Haberl} F.,  {Turolla} R.,  {de Vries} C.~P.,  {Zane} S.,  {Vink} J.,
  {M{\'e}ndez} M.,    {Verbunt} F.,  2006, \aap, 451, L17

\bibitem[\protect\citeauthoryear{{Haberl}, {Zavlin}, {Tr{\"u}mper} \&
  {Burwitz}}{{Haberl} et~al.}{2004}]{2004A&A...419.1077H}
{Haberl} F.,  {Zavlin} V.~E.,  {Tr{\"u}mper} J.,    {Burwitz} V.,  2004, \aap,
  419, 1077

\bibitem[\protect\citeauthoryear{{Hambaryan}, {Neuh{\"a}user}, {Haberl},
  {Hohle} \& {Schwope}}{{Hambaryan} et~al.}{2009}]{2009A&A...497L...9H}
{Hambaryan} V.,  {Neuh{\"a}user} R.,  {Haberl} F.,  {Hohle} M.~M.,    {Schwope}
  A.~D.,  2009, \aap, 497, L9

\bibitem[\protect\citeauthoryear{{Heyl} \& {Kulkarni}}{{Heyl} \&
  {Kulkarni}}{1998}]{1998ApJ...506L..61H}
{Heyl} J.~S.,  {Kulkarni} S.~R.,  1998, \apjl, 506, L61

\bibitem[\protect\citeauthoryear{{Hohle}, {Haberl}, {Vink}, {de Vries} \&
  {Neuh{\"a}user}}{{Hohle} et~al.}{2012}]{2012MNRAS.419.1525H}
{Hohle} M.~M.,  {Haberl} F.,  {Vink} J.,  {de Vries} C.~P.,    {Neuh{\"a}user}
  R.,  2012, \mnras, 419, 1525

\bibitem[\protect\citeauthoryear{{Hohle}, {Haberl}, {Vink}, {Turolla},
  {Hambaryan}, {Zane}, {de Vries} \& {M{\'e}ndez}}{{Hohle}
  et~al.}{2009}]{2009A&A...498..811H}
{Hohle} M.~M.,  {Haberl} F.,  {Vink} J.,  {Turolla} R.,  {Hambaryan} V.,
  {Zane} S.,  {de Vries} C.~P.,    {M{\'e}ndez} M.,  2009, \aap, 498, 811

\bibitem[\protect\citeauthoryear{{Hohle}, {Haberl}, {Vink}, {Turolla}, {Zane},
  {de Vries} \& {M{\'e}ndez}}{{Hohle} et~al.}{2010}]{2010A&A...521A..11H}
{Hohle} M.~M.,  {Haberl} F.,  {Vink} J.,  {Turolla} R.,  {Zane} S.,  {de Vries}
  C.~P.,    {M{\'e}ndez} M.,  2010, \aap, 521, A11+

\bibitem[\protect\citeauthoryear{{Juda}}{{Juda}}{1996}]{1996ChNew...4....9J}
{Juda} M.,  1996, Chandra News, 4, 9

\bibitem[\protect\citeauthoryear{{Kaplan} \& {van Kerkwijk}}{{Kaplan} \& {van
  Kerkwijk}}{2005}]{2005ApJ...628L..45K}
{Kaplan} D.~L.,  {van Kerkwijk} M.~H.,  2005, \apjl, 628, L45

\bibitem[\protect\citeauthoryear{{Kaplan} \& {van Kerkwijk}}{{Kaplan} \& {van
  Kerkwijk}}{2009a}]{2009ApJ...705..798K}
{Kaplan} D.~L.,  {van Kerkwijk} M.~H.,  2009a, \apj, 705, 798

\bibitem[\protect\citeauthoryear{{Kaplan} \& {van Kerkwijk}}{{Kaplan} \& {van
  Kerkwijk}}{2009b}]{2009ApJ...692L..62K}
{Kaplan} D.~L.,  {van Kerkwijk} M.~H.,  2009b, \apjl, 692, L62

\bibitem[\protect\citeauthoryear{{Kaplan}, {van Kerkwijk} \&
  {Anderson}}{{Kaplan} et~al.}{2002}]{2002ApJ...571..447K}
{Kaplan} D.~L.,  {van Kerkwijk} M.~H.,    {Anderson} J.,  2002, \apj, 571, 447

\bibitem[\protect\citeauthoryear{{Kaplan}, {van Kerkwijk} \&
  {Anderson}}{{Kaplan} et~al.}{2007}]{2007ApJ...660.1428K}
{Kaplan} D.~L.,  {van Kerkwijk} M.~H.,    {Anderson} J.,  2007, \apj, 660, 1428

\bibitem[\protect\citeauthoryear{{Kaplan}, {van Kerkwijk}, {Marshall},
  {Jacoby}, {Kulkarni} \& {Frail}}{{Kaplan} et~al.}{2003}]{2003ApJ...590.1008K}
{Kaplan} D.~L.,  {van Kerkwijk} M.~H.,  {Marshall} H.~L.,  {Jacoby} B.~A.,
  {Kulkarni} S.~R.,    {Frail} D.~A.,  2003, \apj, 590, 1008

\bibitem[\protect\citeauthoryear{{Motch} \& {Haberl}}{{Motch} \&
  {Haberl}}{1998}]{1998A&A...333L..59M}
{Motch} C.,  {Haberl} F.,  1998, \aap, 333, L59

\bibitem[\protect\citeauthoryear{{Motch}, {Zavlin} \& {Haberl}}{{Motch}
  et~al.}{2003}]{2003A&A...408..323M}
{Motch} C.,  {Zavlin} V.~E.,    {Haberl} F.,  2003, \aap, 408, 323

\bibitem[\protect\citeauthoryear{{Perna} \& {Pons}}{{Perna} \&
  {Pons}}{2011}]{2011ApJ...727L..51P}
{Perna} R.,  {Pons} J.~A.,  2011, \apjl, 727, L51

\bibitem[\protect\citeauthoryear{{Popov}, {Pons}, {Miralles}, {Boldin} \&
  {Posselt}}{{Popov} et~al.}{2010}]{2010MNRAS.401.2675P}
{Popov} S.~B.,  {Pons} J.~A.,  {Miralles} J.~A.,  {Boldin} P.~A.,    {Posselt}
  B.,  2010, \mnras, 401, 2675

\bibitem[\protect\citeauthoryear{{Rea}, {Esposito}, {Turolla}, {Israel},
  {Zane}, {Stella}, {Mereghetti}, {Tiengo}, {G{\"o}tz}, {G{\"o}{\u g}{\"u}{\c
  s}} \& {Kouveliotou}}{{Rea} et~al.}{2010}]{2010Sci...330..944R}
{Rea} N.,  {Esposito} P.,  {Turolla} R.,  {Israel} G.~L.,  {Zane} S.,  {Stella}
  L.,  {Mereghetti} S.,  {Tiengo} A.,  {G{\"o}tz} D.,  {G{\"o}{\u g}{\"u}{\c
  s}} E.,    {Kouveliotou} C.,  2010, Science, 330, 944

\bibitem[\protect\citeauthoryear{{Str{\"u}der}, {Briel}, {Dennerl}, {Hartmann},
  {Kendziorra}, {Meidinger}, {Pfeffermann} \& {Reppin}}{{Str{\"u}der}
  et~al.}{2001}]{2001A&A...365L..18S}
{Str{\"u}der} L.,  {Briel} U.,  {Dennerl} K.,  {Hartmann} R.,  {Kendziorra} E.,
   {Meidinger} N.,  {Pfeffermann} E.,    {Reppin} C. e.~a.,  2001, \aap, 365,
  L18

\bibitem[\protect\citeauthoryear{{Tetzlaff}, {Neuh{\"a}user}, {Hohle} \&
  {Maciejewski}}{{Tetzlaff} et~al.}{2010}]{2010MNRAS.402.2369T}
{Tetzlaff} N.,  {Neuh{\"a}user} R.,  {Hohle} M.~M.,    {Maciejewski} G.,  2010,
  \mnras, 402, 2369

\bibitem[\protect\citeauthoryear{{Turner}, {Abbey}, {Arnaud}, {Balasini},
  {Barbera}, {Belsole}, {Bennie} \& {Bernard}}{{Turner}
  et~al.}{2001}]{2001A&A...365L..27T}
{Turner} M.~J.~L.,  {Abbey} A.,  {Arnaud} M.,  {Balasini} M.,  {Barbera} M.,
  {Belsole} E.,  {Bennie} P.~J.,    {Bernard} J.~P. e.~a.,  2001, \aap, 365,
  L27

\bibitem[\protect\citeauthoryear{{Turolla}, {Zane}, {Pons}, {Esposito} \&
  {Rea}}{{Turolla} et~al.}{2011}]{2011ApJ...740..105T}
{Turolla} R.,  {Zane} S.,  {Pons} J.~A.,  {Esposito} P.,    {Rea} N.,  2011,
  \apj, 740, 105

\bibitem[\protect\citeauthoryear{{van Kerkwijk} \& {Kaplan}}{{van Kerkwijk} \&
  {Kaplan}}{2008}]{2008ApJ...673L.163V}
{van Kerkwijk} M.~H.,  {Kaplan} D.~L.,  2008, \apjl, 673, L163

\bibitem[\protect\citeauthoryear{{van Kerkwijk}, {Kaplan}, {Pavlov} \&
  {Mori}}{{van Kerkwijk} et~al.}{2007}]{2007ApJ...659L.149V}
{van Kerkwijk} M.~H.,  {Kaplan} D.~L.,  {Pavlov} G.~G.,    {Mori} K.,  2007,
  \apjl, 659, L149

\bibitem[\protect\citeauthoryear{{Vink}, {de Vries}, {M{\'e}ndez} \&
  {Verbunt}}{{Vink} et~al.}{2004}]{2004ApJ...609L..75V}
{Vink} J.,  {de Vries} C.~P.,  {M{\'e}ndez} M.,    {Verbunt} F.,  2004, \apjl,
  609, L75

\bibitem[\protect\citeauthoryear{{Zane}, {Haberl}, {Cropper}, {Zavlin}, {Lumb},
  {Sembay} \& {Motch}}{{Zane} et~al.}{2002}]{2002MNRAS.334..345Z}
{Zane} S.,  {Haberl} F.,  {Cropper} M.,  {Zavlin} V.~E.,  {Lumb} D.,  {Sembay}
  S.,    {Motch} C.,  2002, \mnras, 334, 345

\end{thebibliography}


\end{document}